\documentclass[11pt]{article}

\usepackage{amsmath}
\usepackage{amssymb}
\usepackage{tikz}
\usepackage{pgfplots}
\usepackage{graphicx}
\usepackage{slashbox}
\usepackage{caption}
\usepackage{subcaption}
\usepackage{xcolor}
\allowdisplaybreaks  
\usepackage{cite}
\usepackage{algorithmicx}
\usepackage{algpseudocode}
\usepackage{algorithm}
\usepackage{subcaption}
\usepackage{hyperref}

\allowdisplaybreaks

\newtheorem{definition}{Definition} % definition numbers are dependent on theorem numbers
\newtheorem{theorem}{Theorem} % same for example numbers
\newtheorem{remark}{Remark}
\newtheorem{lemma}{Lemma}
\newtheorem{corollary}{Corollary}
\newtheorem{example}{Example}
\usepackage{authblk}
\usepackage{setspace}
\usepackage{fullpage,etoolbox}
\title{STL Robustness Risk over Discrete-Time Stochastic Processes\thanks{This research was generously  supported by NSF award CPS-2038873, NSF CAREER award ECCS-2045834, and AFOSR grant FA9550-19-1-0265 (Assured Autonomy in Contested Environments).}}
\author{Lars Lindemann}
\author{Nikolai Matni}
\author{George J. Pappas\thanks{The authors are with the Department of Electrical and Systems Engineering, University of Pennsylvania, Philadelphia, PA 19104, USA. Email:  {\tt\small larsl@seas.upenn.edu (L. Lindemann), nmatni@seas.upenn.edu (N. Matni), and pappasg@seas.upenn.edu (G.J. Pappas)}.}}
\affil{Department of Electrical and Systems Engineering, University of Pennsylvania}

\begin{document}

\maketitle

%%%%%%%%%%%%%%%%%%%%%%%%%%%%%%%%%%%%%%%%%%%%%%%%%%%%%%%%%%%%%%%%%%%%%%%%%%%%%%%%
\begin{abstract}
We present a framework to interpret signal temporal logic (STL) formulas over discrete-time stochastic processes in terms of the induced risk. Each realization of a stochastic process either satisfies or violates an STL formula. In fact, we can assign a robustness value to each realization that indicates how robustly this realization satisfies an STL formula. We then define the risk of a stochastic process not satisfying an STL formula robustly, referred to as the \emph{STL robustness risk}. In our definition, we permit general classes of risk measures such as, but not limited to, the conditional value-at-risk. While in general hard to compute, we propose an approximation of the STL robustness risk. This approximation has the desirable property of being an upper bound of the STL robustness risk when the chosen risk measure is monotone, a property satisfied by most  risk measures. Motivated by the interest in data-driven approaches, we present a sampling-based method for estimating the approximate STL robustness risk from data for the value-at-risk. While we consider the value-at-risk, we highlight that such sampling-based methods are  viable for other risk measures.  
\end{abstract}

%%%%%%%%%%%%%%%%%%%%%%%%%%%%%%%%%%%%%%%%%%%%%%%%%%%%%%%%%%%%%%%%%%%%%%%%%%%%%%%%
\section{Introduction}
\label{sec:introduction}

Consider the  scenario in which an autonomous car equipped with noisy sensors navigates through urban traffic. As a consequence of imperfect sensing,  the environment is not perfectly known. Instead, we can describe the scenario  as a stochastic process that models each possible outcome along with the probability of an outcome. In this paper, we are  interested in quantifying the associated risk in such safety-critical systems. In particular, we consider system specifications that are formulated in signal temporal logic (STL) \cite{maler2004monitoring} and, for the first time, propose a systematic way to assess the risk associated with such system specifications when evaluated over discrete-time stochastic processes. 

Signal temporal logic has been introduced as a formalism to express a large variety of complex system specifications. STL particularly allows to express temporal and spatial system properties, e.g., surveillance (``visit regions A, B, and C every $10-60$ sec"), safety (``always between $5-25$ sec stay at least $1$ m away from region D"), and many others. STL specifications are evaluated over deterministic signals and a given signal, for instance the trajectory of a robot, either satisfies or violates the STL specification at hand. Towards quantifying the robustness by which a signal satisfies an STL specification, the authors in \cite{fainekos2009robustness} proposed the robustness degree as a  tube around a nominal signal so that all signals in this tube satisfy (violate) the specification if  the nominal signal satisfies (violates) the specification. In this way, the size of the tube indicates the robustness of the nominal signal with respect to the specification. As the robustness degree is in general hard to calculate, the authors in \cite{fainekos2009robustness}  proposed approximate yet easier to calculate robust semantics. Several  other approximations have appeared such as the space and  time robustness \cite{donze2}, the arithmetic-geometric mean robustness \cite{mehdipour}, the smooth cumulative robustness \cite{haghighi2019control}, averaged STL \cite{akazaki2015time}, or robustness metrics tailored for guiding reinforcement learning  \cite{varnai2020robustness}. Also related is the work \cite{madsen2018metrics} where metrics for STL formulas are presented and \cite{rodionova2016temporal} where a connection with linear time-invariant filtering is made.

The aforementioned works deal with deterministic signals. For stochastic signals, the authors in \cite{tiger2020incremental,li2017stochastic,kyriakis2019specification,SadighRSS16,jha2018safe} propose notions of probabilistic signal temporal logic in which chance constraints are defined over the atomic elements (called predicates) of STL, while the Boolean and temporal operators of STL are not altered. Similarly, notions of risk signal temporal logic have recently appeared in \cite{lindemann2020control} and \cite{safaoui2020control} by defining risk constraints over the atomic elements only. The work in \cite{farahani2018shrinking} considers the probability of an STL specification being satisfied instead of applying chance or risk constraints on the atomic level. More with a control synthesis focus and for the less expressive formalism of linear temporal logic, the authors  in \cite{bharadwaj2018synthesis,vasile2016control,lahijanian2015formal} consider control over belief spaces, while the authors in \cite{guo2018probabilistic} consider probabilistic satisfaction over Markov decision processes. In contrast, in this work we quantify the risk of not satisfying an STL specification robustly.  Probably closest to our paper are  \cite{bartocci2013robustness} and \cite{bartocci2015system} in which the authors present a framework for the robustness of STL under stochastic models. Our work differs from these in several directions. Most importantly, we do not limit our attention to average satisfaction, termed average robustness degree and defined via the distribution of the approximate robustness degree. We instead allow  for general risk measures towards an axiomatic risk theory for temporal logics. We also argue that the STL robustness risk should conceptually be defined differently than the average robustness degree in \cite{bartocci2013robustness} and \cite{bartocci2015system}. We further present an efficient way to reliably estimate the STL robustness risk for the value-at-risk.

The theory of risk has a long history in finance \cite{rockafellar2000optimization,rockafellar2002conditional}. More recently, there has been an interest to also apply such risk measures in robotics and control applications \cite{majumdar2020should}. Risk-aware control and estimation frameworks have recently appeared in \cite{samuelson2018safety,hyeon2020fast,mcgill2019probabilistic,singh2018framework,kalogerias2020better,chapman2019risk,9147792,schuurmans2020learning} using various forms of risk. We remark that these frameworks are orthogonal to our work as they present design tools while we provide a generic framework for quantifying the risk of complex system specifications expressed in STL. We hope that such quantification will be useful to guide the design and analysis process in the future. 

In this paper, we consider signal temporal logic specifications interpreted over discrete-time stochastic processes. Our contributions can be summarized as follows:
\begin{enumerate}
    \item We show that the semantics, the robust semantics, and the robustness degree of STL are measurable functions so that these functions are well-defined and have a probability distribution.
    \item We define the risk of a discrete-time stochastic process not satisfying an STL specification robustly and refer to this definition as the ``STL robustness risk''.
    \item We argue that the robustness risk is in general hard to calculate and propose an approximation of the robustness risk that has the desireable property of being an upper bound of the STL robustness risk, i.e., more risk averse, if the risk measure is monotone.
    \item We present a sampling-based estimate of the approximate robustness risk for the value-at-risk. We show that this estimate is an upper bound of the approximate robustness risk with high probability. We thereby establish an interesting connection between data-driven design approaches and the risk of an STL specification. 
\end{enumerate}

In Section \ref{sec:backgound}, we present background on signal temporal logic, stochastic processes, and risk measures. In Section \ref{risskk}, we define the STL robustness risk, while we show in Section \ref{comppp} how the approximate robustness risk can be obtained via a  sampling-based method for the case of the value-at-risk. A case study is presented in Section \ref{sec:simulations} followed by conclusions in Section \ref{sec:conclusion}. All proofs of the presented theorems can be found in the appendix.

\section{Background}
\label{sec:backgound}

True and false are encoded as $\top:=1$ and $\bot:=-1$, respectively, with the set $\mathbb{B}:=\{\top,\bot\}$. Let $\mathbb{R}$ and $\mathbb{N}$ be the set of real and natural numbers. Let $\overline{\mathbb{R}}:=\mathbb{R}\cup\{\infty,-\infty\}$ be the set of extended real numbers. Also let $\mathbb{R}_{\ge 0}$ be the set of non-negative real numbers and $\mathbb{R}^n$ be the real $n$-dimensional vector space. For a metric space $(S,d)$, a point $s\in S$, and a nonempty set $S'\subseteq S$, let $\bar{d}(s,S'):=\inf_{s'\in S'} d(s,s')$ be the distance of $s$ to $S'$. It  holds that the function $\bar{d}(s,S')$ is continuous in $s$ \cite[Chapter 3]{munkres1975prentice}.  We use the extended definition of the supremum and infimum operators, i.e., the supremum of the empty set is the smallest element of the domain and the infimum of the empty set is the largest element of the domain.  For  $t\in\mathbb{R}$ and $I\subseteq\mathbb{R}$, let  $t\oplus I$ and $t\ominus I$ denote the Minkowski sum and the Minkowski difference of $t$ and $I$, respectively. For  $a,b\in\mathbb{R}$, let
\begin{align*}
\mathbb{I}(a\le b):=\begin{cases}
1 &\text{if } a\le b\\
0 &\text{otherwise}
\end{cases}
\end{align*} 
be the indicator function.  Let $\mathfrak{F}(T,S)$ denote the set of all measurable functions mapping from the domain $T$ into the domain $S$, i.e., $f\in \mathfrak{F}(T,S)$ is a function $f:T\to S$.

\subsection{Signal Temporal Logic}
\label{sec:STL}
Signal temporal logic \cite{maler2004monitoring} is based on deterministic signals $x:T\to\mathbb{R}^n$ where $T:=\mathbb{N}$ is assumed throughout the paper. The atomic elements of STL are predicates that are functions $\mu:\mathbb{R}^n\to\mathbb{B}$. Let now $M$ be a set of such predicates $\mu$ and let us associate an observation map $O^\mu\subseteq \mathbb{R}^n$ with $\mu$. The observation map $O^\mu$ indicates regions within the state space where $\mu$ is true, i.e.,
\begin{align*}
O^\mu:=\mu^{-1}(\top)
\end{align*}
where $\mu^{-1}(\top)$ denotes the inverse image of $\top$ under $\mu$. We assume throughout the paper that the sets $O^\mu$ and $O^{\neg\mu}$ are non-empty and measurable for any $\mu\in M$, i.e., $O^\mu$ and $O^{\neg\mu}$  are elements of the Borel $\sigma$-algebra $\mathcal{B}^n$ of $\mathbb{R}^n$. 
\begin{remark}
For convenience, the predicate $\mu$ is often defined via a predicate function $h:\mathbb{R}^n\to\mathbb{R}$ so that
\begin{align*}
\mu(\zeta):=\begin{cases}
\top & \text{if } h(\zeta)\ge 0\\
\bot &\text{otherwise}
\end{cases}
\end{align*}
for $\zeta\in\mathbb{R}^n$. In this case, we have  $O^\mu=\{\zeta\in\mathbb{R}^n|h(\zeta)\ge 0\}$.
\end{remark}

For $\mu\in M$, the syntax of STL, also referred to as the grammar of STL, is defined as 
\begin{align}\label{eq:full_STL}
\phi \; ::= \; \top \; | \; \mu \; | \;  \neg \phi \; | \; \phi' \wedge \phi'' \; | \; \phi'  U_I \phi'' \; | \; \phi' \underline{U}_I \phi'' \,
\end{align}
where $\phi'$ and $\phi''$ are STL formulas and where $U_I$ is the future until operator with $I\subseteq \mathbb{R}_{\ge 0}$, while $\underline{U}_I$ is the past until-operator. The operators $\neg$ and $\wedge$ encode negations and conjunctions. Also define the set of operators
\begin{align*}
\phi' \vee \phi''&:=\neg(\neg\phi' \wedge \neg\phi'') &\text{ (disjunction operator)},\\
F_I\phi&:=\top U_I \phi &\text{ (future eventually operator)},\\
\underline{F}_I\phi&:=\top \underline{U}_I \phi &\text{ (past eventually operator)},\\
G_I\phi&:=\neg F_I\neg \phi &\text{ (future always operator)},\\
\underline{G}_I\phi&:=\neg \underline{F}_I\neg \phi &\text{ (past always operator).}
\end{align*}

\subsubsection{Semantics} We can now  give an STL formula $\phi$ as in \eqref{eq:full_STL} a meaning  by defining the satisfaction function $\beta^\phi:\mathfrak{F}(T,\mathbb{R}^n)\times T \to \mathbb{B}$. In particular, $\beta^\phi(x,t)=\top$ indicates that the signal $x$ satisfies the formula $\phi$ at time $t$, while $\beta^\phi(x,t)=\bot$ indicates that $x$ does not satisfy $\phi$ at time $t$. For a formal definition of $\beta^\phi(x,t)$, we refer to Definition \ref{def:qualitative_semantics} in Appendix \ref{app:STL}. An STL formula $\phi$ is said to be satisfiable if $\exists x\in \mathfrak{F}(T,\mathbb{R}^n)$ such that $\beta^\phi(x,0)=\top$. The following example is used as a running example throughout the paper. 
\begin{example}\label{ex1}
Consider a scenario in which a robot operates in a hospital environment. The robot needs to perform two time-critical sequential delivery tasks in regions $A$ and $B$ while avoiding areas $C$ and $D$ in which potentially humans operate.  In particular, we consider the STL formula 
\begin{align}\label{ex:1_formula}
    \phi:=G_{[0,3]} (\neg \mu_{C} \wedge \neg\mu_{D}) \wedge F_{[1,2]}(\mu_{A} \wedge F_{[0,1]}\mu_{B}).
\end{align}
To define  $\mu_A$, $\mu_B$, $\mu_C$, and $\mu_D$, let $a$, $b$, $c$, and $d$ denote the midpoints of the regions $A$, $B$, $C$, and $D$ as
\begin{align*}
    a:=\begin{bmatrix}
        4 & 5
    \end{bmatrix}^T
     \hspace{0.5cm} b:=\begin{bmatrix}
        7 & 2
    \end{bmatrix}^T \hspace{0.5cm}
     c:=\begin{bmatrix}
        2 & 3
    \end{bmatrix}^T
     \hspace{0.5cm}d:=\begin{bmatrix}
        6 & 4
    \end{bmatrix}^T.
\end{align*}
Also let the state $x(t)\in\mathbb{R}^{10}$ at time $t$ be defined as
\begin{align*}
    x(t):=\begin{bmatrix}r(t) & a & b& c& d \end{bmatrix}^T
\end{align*} 
where $r(t)$ is the robot position at time $t$. The predicates $\mu_A$, $\mu_B$, $\mu_C$, and $\mu_D$ are now described by the observation maps
\begin{align*}
    O^{\mu_A}&:=\{x\in\mathbb{R}^{10}|\|r-a\|_\infty\le 0.5\},\\
    O^{\mu_B}&:=\{x\in\mathbb{R}^{10}|\|r-b\|_2\le 0.7\},\\
    O^{\mu_C}&:=\{x\in\mathbb{R}^{10}|\|r-c\|_\infty\le 0.5\},\\
    O^{\mu_D}&:=\{x\in\mathbb{R}^{10}|\|r-d\|_2\le 0.7\}.
\end{align*}
where $\|\cdot\|_2$ is the Euclidean and $\|\cdot\|_\infty$ is the infinity norm. In Fig. \ref{ex:1_figure}, six different robot trajectories $r_1$-$r_6$ are displayed. We omit the exact timings associated with $r_1$-$r_6$ in Fig. \ref{ex:1_figure} for readability. However, it can be seen that the signal $x_1$ that corresponds to $r_1$ violates $\phi$ as the region $D$ is entered, while $x_2$-$x_6$ satisfy $\phi$. In other words, we have $\beta^\phi(x_1,0)=\bot$ and $\beta^\phi(x_j,0)=\top$ for all $j\in\{2,\hdots,6\}$. 
\begin{figure}
\centering
\includegraphics[scale=0.5]{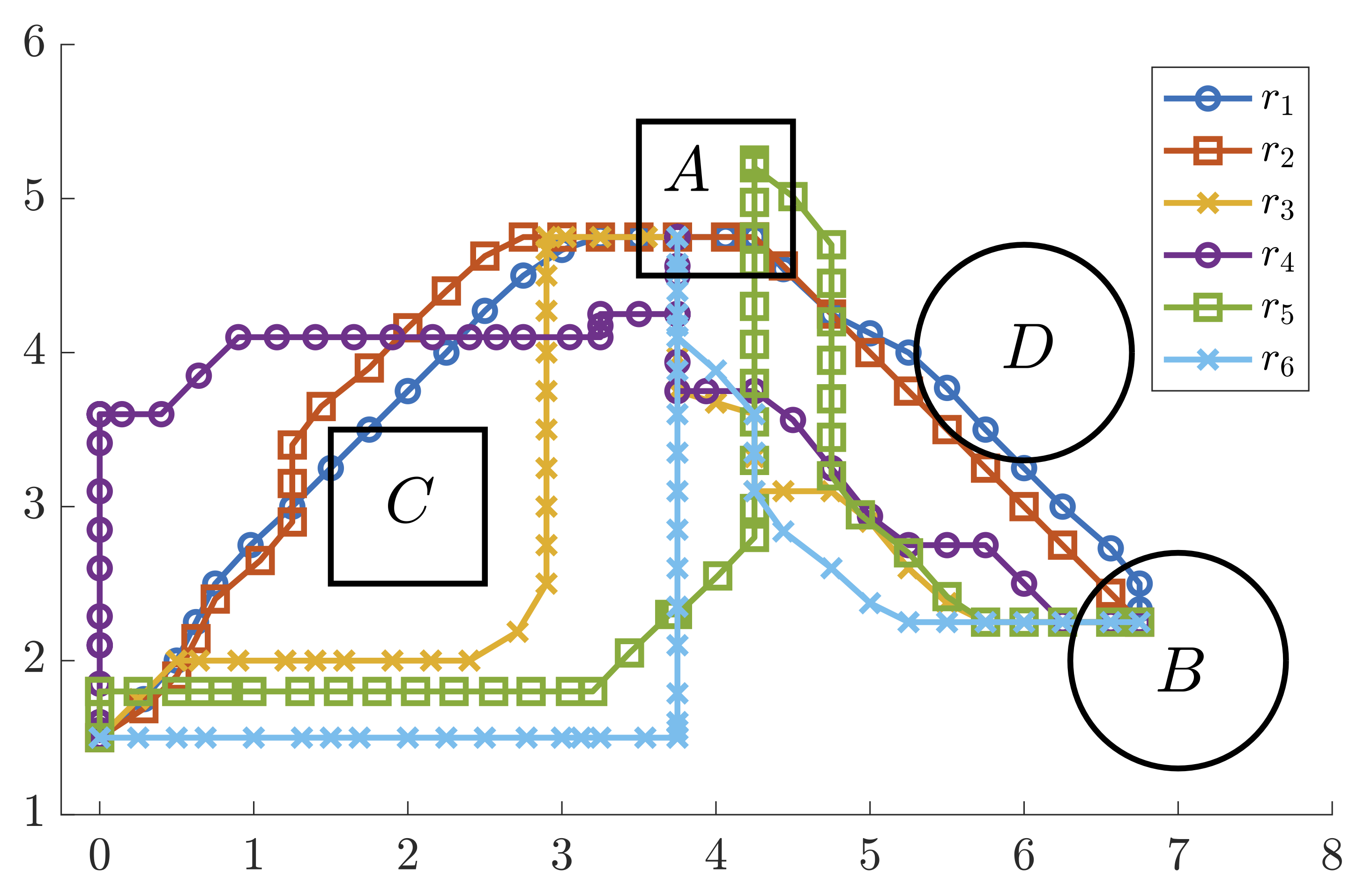}
\caption{Shown are six robot trajectories $r_1$-$r_6$ along with four regions $A$, $B$, $C$, and $D$. The specification given in \eqref{ex:1_formula} is violated by $r_1$ and satisfied by $r_2$-$r_6$. Furthermore, $r_2$ only marginally satisfies  $\phi$, while $r_3$-$r_6$ satisfy $\phi$ robustly.}
\label{ex:1_figure}
\end{figure}
\end{example}
%\begin{remark}\label{remmm}
%	Definition \ref{def:qualitative_semantics} considers  the strict non-matching version of the until operator. In some situations, we will instead replace the  strict non-matching version  by the non-strict matching version of the until operator. This results in replacing the open time intervals $(t,t'')$ in Definition \ref{def:qualitative_semantics} by the closed time intervals $[t,t'']$ (see \cite{fainekos2009robustness} for more details)
%\end{remark}
	
\subsubsection{Robustness}	One may now be interested in more information than just whether or not the signal $x$ satisfies the STL formula $\phi$ at time $t$ and consider  the quality of satisfaction. Therefore, one can look at the robustness by which a signal $x$ satisfies the STL formula $\phi$ at time $t$. For this purpose, the \emph{robustness degree} has been introduced in \cite[Definition 7]{fainekos2009robustness}.	 Let us define the set of  signals that satisfy $\phi$ at time $t$ as
	\begin{align*}
	\mathcal{L}^\phi(t):=\{x\in \mathfrak{F}(T,\mathbb{R}^n)|\beta^\phi(x,t)=\top\}.
	\end{align*} 
	 Let us also define the signal metric
	\begin{align*}
	\kappa(x,x^*):=\sup_{t\in T} d\big(x(t),x^*(t)\big)
	\end{align*}
	where $d:\mathbb{R}^n\times\mathbb{R}^n\to\overline{\mathbb{R}}$ is a vector metric, e.g., the Euclidean distance between two points in $\mathbb{R}^n$. Note that $\kappa(x,x^*)$ is the $L_\infty$ norm of the signal $x-x^*$. The distance of $x$ to the set $\mathcal{L}^\phi(t)$ is then defined via the metric $\kappa$ as 
	\begin{align*}
	\text{dist}^\phi(x,t)=\bar{\kappa}\big(x,\text{cl}(\mathcal{L}^\phi(t))\big):=\inf_{x^*\in \text{cl}(\mathcal{L}^\phi(t))}\kappa(x,x^*),
	\end{align*} 
	where $\text{cl}(\mathcal{L}^\phi(t))$ denotes the closure of $\mathcal{L}^\phi(t)$. The  robustness degree is now given in Definition \ref{def:rd_cont}.
	
	\begin{definition}[Robustness Degree]\label{def:rd_cont}
		Given an STL formula $\phi$ and a signal $x\in\mathfrak{F}(T,\mathbb{R}^n)$, the robustness degree at time $t$ is defined as \cite[Definition~7]{fainekos2009robustness}:
		\begin{align*}
		\mathcal{RD}^\phi(x,t):=
		\begin{cases}
		\text{dist}^{\neg\phi}(x,t) \text{ if } x\in \mathcal{L}^\phi(t)\\
		-\text{dist}^\phi(x,t) \text{ if } x\notin \mathcal{L}^\phi(t).
		\end{cases}
		\end{align*}
	\end{definition}

Intuitively, the robustness degree then tells us how much the signal $x$ can be perturbed by additive noise before changing the Boolean truth value of the specification $\phi$. In other words, if $|\mathcal{RD}^\phi(x,t)|\neq 0$ and $x\in {\mathcal{L}}^\phi(t)$, it follows that all signals $x^*\in\mathfrak{F}(T,\mathbb{R}^n)$ that are such that $\kappa(x,x^*)< |\mathcal{RD}^\phi(x,t)|$ satisfy $x^*\in {\mathcal{L}}^\phi(t)$. 

The robustness degree is a \emph{robust neighborhood}. A robust neighborhood of $x$ is a tube of diameter $\epsilon\ge 0$ around $x$ so that for all $x^*$ in this tube we have $\beta^\phi(x,t)=\beta^\phi(x^*,t)$. Specifically, for $\epsilon\ge 0$ and $x:T\to\mathbb{R}^n$ with $x\in \mathcal{L}^\phi(t)$, a set $\{x'\in\mathfrak{F}(T,\mathbb{R}^n)|\kappa(x,x')< \epsilon\}$ is a robust neighborhood if $x^*\in\{x'\in\mathfrak{F}(T,\mathbb{R}^n)|\kappa(x,x')< \epsilon\}$ implies $x^*\in \mathcal{L}^\phi(t)$. 

\subsubsection{Robust Semantics} Note that it is in general difficult to calculate the robustness degree $\mathcal{RD}^\phi(x,t)$ as the set $\mathcal{L}^\phi(t)$ is hard to calculate. The  authors in \cite{fainekos2009robustness} introduce the \emph{robust semantics} $\rho^\phi:\mathfrak{F}(T,\mathbb{R}^n)\times T\to \mathbb{R}$ as an alternative way of finding a robust neighborhood. 

\begin{definition}[STL Robust Semantics]\label{def:quantitative_semantics}
For a signal $x:T\to\mathbb{R}^n$, the robust semantics $\rho^\phi(x,t)$  of an STL formula $\phi$ are inductively defined as 
	\begin{align*}
	\rho^{\top}(x,t)& := \infty,\\
	\rho^{\mu}(x,t)& := \begin{cases} \text{dist}^{\neg\mu}(x,t) &\text{if } x\in \mathcal{L}^\mu(t)\\
	-\text{dist}^{\mu}(x,t) &\text{otherwise,}
	\end{cases}\\
	\rho^{\neg\phi}(x,t) &:= 	-\rho^{\phi}(x,t),\\
	\rho^{\phi' \wedge \phi''}(x,t) &:= 	\min(\rho^{\phi'}(x,t),\rho^{\phi''}(x,t)),\\
	%	\rho^{\phi' \vee \phi''}(x,t) &:= 	\max(\rho^{\phi'}(x,t),\rho^{\phi''}(x,t)),\\
	\rho^{\phi' U_I \phi''}(x,t) &:= \underset{t''\in (t\oplus I)\cap T}{\text{sup}}  \Big(\min\big(\rho^{\phi''}(x,t''),\underset{t'\in (t,t'')\cap T}{\text{inf}}\rho^{\phi'}(x,t') \big)\Big), \\
	\rho^{\phi' \underline{U}_I \phi''}(x,t) &:= \underset{t''\in (t\ominus I)\cap T}{\text{sup}} \Big( \min\big(\rho^{\phi''}(x,t''),  \underset{t'\in (t,t'')\cap T}{\text{inf}}\rho^{\phi'}(x,t') \big)\Big).
	%	\rho^{G_I \phi}(x,t) &:= \underset{t'\in t\oplus I}{\text{inf}}\rho^{\phi}(x,t'),\\
	%	\rho^{\underline{G}_I \phi}(x,t) &:= \underset{t'\in t\ominus I}{\text{inf}}\rho^{\phi}(x,t'),\\
	%	\rho^{F_I \phi}(x,t) &:= \underset{t'\in t\oplus I}{\text{sup}}\rho^{\phi}(x,t'),\\
	%	\rho^{\underline{F}_I \phi}(x,t) &:= \underset{t'\in t\ominus I}{\text{sup}}\rho^{\phi}(x,t').
	\end{align*}
\end{definition}
Importantly, it was shown in \cite[Theorem 28]{fainekos2009robustness} that 
\begin{align}\label{underapprox}
-\text{dist}^\phi(x,t)\le \rho^{\phi}(x,t)\le \text{dist}^{\neg\phi}(x,t).
\end{align} 
In other words, it holds that
\begin{align*}
\begin{cases}
0\le \rho^\phi(x,t)\le \text{dist}^{\neg\phi}(x,t) \text{ if } x\in \mathcal{L}^\phi(t) \\
-\text{dist}^\phi(x,t)\le \rho^\phi(x,t)\le 0 \text{ if } x\notin \mathcal{L}^\phi(t)
\end{cases} 
\end{align*}
so that $|\rho^\phi(x,t)|\le |\mathcal{RD}^\phi(x,t)|$. The robust semantics $\rho^\phi(x,t)$ hence provide a more tractable under-approximation of the robustness degree $\mathcal{RD}^\phi(x,t)$. The robust semantics are sound in the following sense \cite[Proposition 30]{fainekos2009robustness}:
\begin{align*}
\beta^\phi(x,t)=\top  \text{ if }\rho^\phi(x,t)> 0,\\
\beta^\phi(x,t)=\bot \text{ if }\rho^\phi(x,t)< 0.
\end{align*} 
This result  allows to use the robust semantics when reasoning over satisfaction of an STL formula $\phi$. 
\begin{example}\label{ex2}
For Example \ref{ex1} and the trajectories shown in Fig. \ref{ex:1_figure}, we obtain $\rho^\phi(x_1,0)=-0.15$, $\rho^\phi(x_2,0)=0.01$, and $\rho^\phi(x_j,0)=0.25$ for all $j\in\{3,\hdots,6\}$ when choosing $d(\cdot)$ as the Euclidean distance. The reason for $x_1$ having negative robustness lies in $r_1$ intersecting with the region $D$. Marginal robustness of $x_2$ is explained as $r_2$ only marginally avoids the region $D$ while all other trajectories avoid the region $D$ robustly.
\end{example}

\subsection{Random Variables and Stochastic Processes}
\label{sec:stoch}
Instead of interpreting an STL specifications $\phi$ over deterministic signals, we will interpret $\phi$ over stochastic processes. Consider therefore the \emph{probability space} $(\Omega,\mathcal{F},P)$  where $\Omega$ is the sample space, $\mathcal{F}$ is a $\sigma$-algebra of $\Omega$, and $P:\mathcal{F}\to[0,1]$ is a probability measure. More intuitively, an element in  $\Omega$ is an \emph{outcome} of an experiment, while an element in $\mathcal{F}$ is an \emph{event} that consists of one or more outcomes whose probabilities can be measured by the probability measure $P$. 

\subsubsection{Random Variables} 

Let $Z$ denote a real-valued \emph{random vector}, i.e., a measurable function $Z:\Omega\to\mathbb{R}^n$.\footnote{More precisely, we have $Z:\Omega\times\mathcal{F}\to\mathbb{R}^n\times\mathcal{B}^n$ where $\mathcal{B}^n$ is the Borel $\sigma$-algebra of $\mathbb{R}^n$, i.e.,  $Z$ maps a \emph{measurable space} to yet another measurable space. For convenience, this more involved notation is, however, omitted.} When $n=1$, we  say $Z$ is a \emph{random variable}. We refer to $Z(\omega)$ as a realization of the random vector $Z$ where $\omega\in\Omega$.    Since $Z$ is a measurable function, a distribution can be assigned to  $Z$ and a cumulative distribution function (CDF) $F_Z(z)$ can be defined for $Z$ (see Appendix \ref{app:ran_var}). 

Given a random vector $Z$, we can derive other random variables that we call \emph{derived random variables}. Assume for instance a measurable function $g: \mathbb{R}^n \to \mathbb{R}$ and notice that $G:\Omega\to\mathbb{R}$ with  $G(\omega):=g(Z(\omega))$ becomes yet another random variable since function composition preserves measureability. See \cite{durrett2019probability} for a more detailed discussion.

\subsubsection{Stochastic Processes} A \emph{stochastic process} is a function of the variables $\omega\in\Omega$ and $t\in T$ where $T$ is the time domain. Recall that the time domain is discrete, i.e., $T:=\mathbb{N}$, so that we consider discrete-time stochastic processes. This assumption  is made for simplicity. The presented results  carry over, with some modifications, to the continuous-time case that we defer to another paper.  A stochastic process is now a function $X:T\times \Omega \to \mathbb{R}^n$ where  $X(t,\cdot)$ is a random vector for each fixed $t\in T$. A stochastic process can   be viewed as a collection of random vectors $\{X(t,\cdot)|t\in T\}$  that are defined on a common probability space $(\Omega,\mathcal{F},P)$ and that are indexed by $T$.  For a fixed $\omega\in\Omega$, the function $X(\cdot,\omega)$ is a \emph{realization} of the stochastic process.  Another equivalent definition is that a stochastic process is a collection of deterministic functions of time 
 \begin{align}\label{def:stoch}
\{X(\cdot,\omega)|\omega\in \Omega\}
 \end{align} 
 that are indexed by $\Omega$. While the former definition is intuitive, the latter allows to define a \emph{random function} mapping from the sample space $\Omega$ into the space of functions $\mathfrak{F}(T,\mathbb{R}^n)$. 
 
 %We refer to $X$ as continuous-space stochastic process when the CDF of $X(t,\cdot)$ is continuous and as discrete-space stochastic process otherwise.  

\subsection{Risk Measures}
\label{sec:risk}
A \emph{risk measure} is a function $R:\mathfrak{F}(\Omega,\mathbb{R})\to \mathbb{R}$ that maps from the set of real-valued random variables to the real numbers. In particular, we refer to the input of a risk measure $R$ as the \emph{cost random variable} since typically a cost is associated with the input of $R$. Risk measures hence allow for a risk assessment in terms of such cost random variables. Commonly used risk measures are the expected value, the variance, or the conditional value-at-risk \cite{rockafellar2000optimization}. A particular property of $R$ that we need in this paper is monotonicity. For two cost random variables $Z,Z'\in \mathfrak{F}(\Omega,\mathbb{R})$, the risk measure $R$ is monotone if $Z(\omega) \leq Z'(\omega)$ for all $\omega\in\Omega$ implies that $R(Z) \le R(Z')$. 
\begin{remark}
In Appendix \ref{app:risk}, we summarize other desireable properties of $R$ such as translation invariance, positive homogeneity, subadditivity, commotone additivity, and law invariance. We also provide a summary of existing risk measures. We  emphasize that our presented method is compatible with any of these risk measures as long as they are  monotone. 
\end{remark}

\section{Risk of STL Specifications}
\label{risskk}

\begin{table*}
	\centering
	\begin{tabular}{ |l|l| } 
		\hline
		\footnotesize{Symbol} & \footnotesize{Meaning}  \\
		\hline
		\footnotesize{$x$} & \footnotesize{Deterministic signal $x:T\to\mathbb{R}^n$} \\
		\footnotesize{$\mathfrak{F}(T,\mathbb{R}^n)$} & \footnotesize{Set of all measurable functions mapping from $T$ to $\mathbb{R}^n$} \\ 
		\footnotesize{$\beta^\phi(x,t)$} & \footnotesize{Boolean semantics $\beta^\phi:\mathfrak{F}(T,\mathbb{R}^n)\times T \to \mathbb{B}$ of an STL formula $\phi$} \\ 
		\footnotesize{$\mathcal{L}^\phi(t)$} & \footnotesize{Set of deterministic signals $x$ that satisfy $\phi$ at time $t$} \\ 
		\footnotesize{$\text{dist}^\phi(x,t)$} & \footnotesize{Distance of the signal $x$ to the set $\mathcal{L}^\phi(t)$} \\ 
		\footnotesize{$\mathcal{RD}^\phi(x,t)$} & \footnotesize{Robustness degree $\mathcal{RD}^\phi:\mathfrak{F}(T,\mathbb{R}^n)\times T \to \mathbb{R}$ of an STL formula $\phi$} \\ 
		\footnotesize{$\rho^\phi(x,t)$} & \footnotesize{Robust semantics $\rho^\phi:\mathfrak{F}(T,\mathbb{R}^n)\times T \to \mathbb{R}$ of an STL formula $\phi$} \\ 
		%\footnotesize{$\tau^\phi(x,t)$} & \footnotesize{Robust semantics (in time) $\tau^\phi:\mathbb{R}^n\times T \to \mathbb{R}$ of an STL formula $\phi$ at time $t$} \\ 
		\footnotesize{$X$} & \footnotesize{Stochastic Process $X:T\times\Omega\to\mathbb{R}^n$} \\ 
		%\footnotesize{$R(-\mathcal{RD}^\phi(X_t,t))$} & \footnotesize{Risk (in space) of the stochastic process $X_t$ violating an STL formula $\phi$ at time $t$ - \textbf{Alternative 1}} \\ 
		\footnotesize{$R(-\text{dist}^{\neg\phi}(X,t))$} & \footnotesize{The STL robustness risk, i.e., the risk of the stochastic process $X$ not satisfying}\\
		 & \footnotesize{the STL formula $\phi$ robustly  at time $t$ } \\ 
		\footnotesize{$R(-\rho^{\phi}(X,t))$} & \footnotesize{The approximate STL robustness risk} \\ 
		%\footnotesize{$R(-\tau^{\phi}(X_t,t))$} & \footnotesize{Risk (in time) of the stochastic process $X_t$ violating an STL formula $\phi$ at time $t$} \\ 
		\hline
	\end{tabular}
	\caption{Summary of robustness and risk notions for signal temporal logic.}
	\label{tab:1}
\end{table*}

While an STL formula $\phi$ as defined in Section \ref{sec:STL} is defined over deterministic signals $x$, we will interpret $\phi$ over a stochastic process $X$ as defined in Section \ref{sec:stoch}.

For a particular realization $X(\cdot,\omega)$ of the stochastic process $X$, note that we can evaluate whether or not $X(\cdot,\omega)$ satisfies $\phi$. For the stochastic process $X$, however, it is not clear how to interpret the satisfaction of $\phi$ by $X$. In fact, some realizations of $X$ may satisfy $\phi$ while some other realizations of $X$ may violate $\phi$. To bridge this gap, we us  risk measures as introduced in Section \ref{sec:risk} to argue about the risk of the stochastic process $X$ not satisfying the specification $\phi$.

Before going into the main parts of this paper, we remark that  all important symbols that have been or will be introduced are summarized in Table \ref{tab:1}.

%For  ease of notation, let us equivalently refer to the stochastic process of $X$ as the function $X_t:\Omega\to\mathfrak{F}(T,\mathbb{R}^n)$ with
%\begin{align*}
%X_t:=X(\cdot,\omega)
%\end{align*}  
%so that $X_t(\omega')(t')=X(t',\omega')$ for $t'\in T$ and $\omega'\in\Omega$. 

\subsection{Measurability of STL Semantics and Robustness Degree}

Note that the semantics and the robust semantics as well as the robustness degree become stochastic entities when evaluated over a stochastic process $X$, i.e., the functions $\beta^\phi(X,t)$, $\rho^\phi(X,t)$, and $RD^\phi(X,t)$  become stochastic entities.  We first provide conditions under which  $\beta^\phi(X,t)$ and $\rho^\phi(X,t)$ become (derived) random variables, which boils down to showing that  $\beta^\phi(X(\cdot,\omega),t)$ and $\rho^\phi(X(\cdot,\omega),t)$ are measurable in $\omega$ for a fixed $t\in T$. 
\begin{theorem}\label{thm:1}
	Let $X$ be a discrete-time stochastic process and let $\phi$ be an STL specification. Then $\beta^\phi(X(\cdot,\omega),t)$ and $\rho^\phi(X(\cdot,\omega),t)$  are measurable in $\omega$ for a fixed $t\in T$ so that  $\beta^\phi(X,t)$ and $\rho^\phi(X,t)$ are random variables.
\end{theorem}

%In the case of a continuous-time stochastic process, note that the supremum and infimum operators within the semantics of $\phi' U_I \phi''$ and $\phi' \underline{U}_I \phi''$ are evaluated over continuous time and hence over uncountable sets. This means that $\beta^\phi(X(\cdot,\omega),t)$ and $\rho^\phi(X(\cdot,\omega),t)$ may in general not be measurable in $\omega$ for a fixed $t\in T$. We next state under which conditions measurability of $\beta^\phi(X(\cdot,\omega),t)$ and $\rho^\phi(X(\cdot,\omega),t)$ hold in continuous time.
%\begin{theorem}\label{thm:2}
%	Let $X$ be a continuous-time stochastic process, i.e., $T=\mathbb{R}$, with $X(\cdot,\omega)$ being a continuous function and let $\phi$ be an STL specification. Assume that the strict non-matching version of the until operator is replaced by the non-strict matching version (see Remark \ref{remmm}). Assume  that  the STL formula $\phi$ is such that all temporal intervals $I$ in $\phi$ are compact. Then $\beta^\phi(X(\cdot,\omega),t)$ and $\rho^\phi(X(\cdot,\omega),t)$  are measurable in $\omega$ for a fixed $t\in T$ so that  $\beta^\phi(X,t)$ and $\rho^\phi(X,t)$ are random variables.
%\end{theorem}

By Theorem \ref{thm:1}, the probabilities
$P(\beta^{\phi}(X,t)\in B)$ and $P(\rho^{\phi}(X,t)\in B)$\footnote{We use the shorthand notations $P(\beta^{\phi}(X,t)\in B)$ and $P(\rho^{\phi}(X,t)\in B)$ instead of the more complex notations $P(\{\omega\in\Omega|\beta^{\phi}(X(\cdot,\omega),t)\in B\})$ and $P(\{\omega\in\Omega|\rho^{\phi}(X(\cdot,\omega),t)\in B\})$, respectively.} are well defined for measurable sets $B$ from the corresponding measurable space. 

We next show measurability of the distance function $\text{dist}^\phi(X(\cdot,\omega),t)$ and the robustness degree $\mathcal{RD}^\phi(X(\cdot,\omega),t)$.
\begin{theorem}\label{thm:2_}
	Let $X$ be a discrete-time stochastic process and let $\phi$ be an STL specification. Then $\text{dist}^\phi(X(\cdot,\omega),t)$ and $\mathcal{RD}^\phi(X(\cdot,\omega),t)$ are measurable in $\omega$ for a fixed $t\in T$ so that $\text{dist}^\phi(X,t)$ and $\mathcal{RD}^\phi(X,t)$ are random variables.
\end{theorem}

%\re{Note, however, that measurability of the robustness degree $RD^\phi(X(\cdot,\omega),t)$ for a fixed $t\in T$ is not as straightforward due to the infimum over the set $\text{cl}(\mathcal{L}^\phi(t))$ in the definition of the distance function $\text{dist}^\phi(x,t)$. We leave this for future work and assume that $RD^\phi(X(\cdot,\omega),t)$ is measurable without much loss of generality since we will use $\beta^\phi(X,t)$ and $\rho^\phi(X,t)$ that are more tractable to compute.}

\subsection{The STL Robustness Risk}

Towards defining the risk of  not satisfying a specification $\phi$,  note that the expression $R(\beta^\phi(X,t)=\bot)$ is not well defined as opposed to $P(\beta^\phi(X,t)=\bot)$  that indicates the probability of not satisfying $\phi$. The reason for this is that the function $R$ takes a real-valued cost random variable as its input. We can instead evaluate $R(-\beta^\phi(X,t))$,  but not much information will be gained  due to the binary encoding  of the STL semantics $\beta^\phi(X,t)$.

\subsubsection{The risk of not satisfying $\phi$ robustly} Instead, we will define  the risk of the stochastic process $X$ not satisfying $\phi$ robustly by considering $\text{dist}^{\neg\phi}(X,t)$. As shown, $\text{dist}^{\neg\phi}(X,t)$ is a random variable indicating the distance between realizations of the stochastic process  and the set of signals $\mathcal{L}^{\neg\phi}(t)$ that violate $\phi$. We refer to the following definition as the \emph{STL robustness risk} for brevity. 
 \begin{definition}[STL Robustness Risk]\label{def:rr}
 	Given an STL formula $\phi$ and a stochastic process $X:T\times\Omega\to\mathbb{R}^n$, the risk of $X$ not satisfying $\phi$ robustly at time $t$ is defined as 
 	\begin{align*}
 	R(-\text{dist}^{\neg\phi}(X,t)).
 	\end{align*}  
 \end{definition}

\begin{figure}
	\centering
	\includegraphics[scale=0.35]{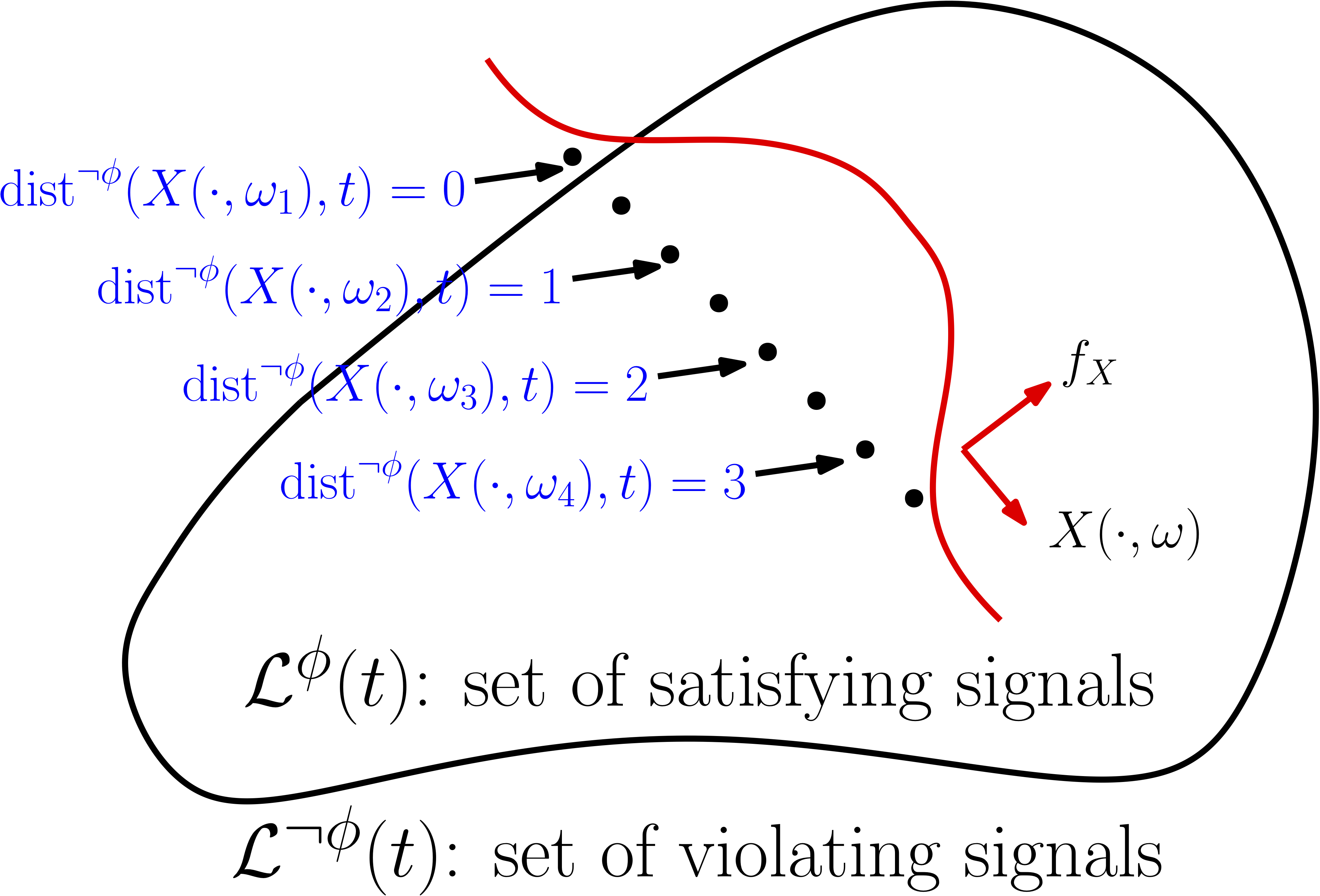}
	\caption{Illustration of $\text{dist}^{\neg\phi}(X(\cdot,\omega_i),t)$, i.e., the distance between realizations $X(\cdot,\omega_i)$ and the set $\mathcal{L}^{\neg\phi}(t)$. Note that the figure shows the function space $\mathfrak{F}(T,\mathbb{R}^n)$ and not $\mathbb{R}^n$.}\label{fig:RD}
\end{figure}

Fig. \ref{fig:RD}  illustrates the idea underlying Definition \ref{def:rr} and shows $\text{dist}^{\neg\phi}(X(\cdot,\omega_i),t)$ for realizations $X(\cdot,\omega_i)$ of the stochastic process $X$ where $\omega_i\in\Omega$, i.e., $\text{dist}^{\neg\phi}(X(\cdot,\omega_i),t)$ is the distance between the realization $X(\cdot,\omega_i)$  and the set $\mathcal{L}^{\neg\phi}(t)$.  Positive values of $\text{dist}^{\neg\phi}(X(\cdot,\omega_i),t)$ indicate that the realization $X(\cdot,\omega_i)$ satisfies $\phi$ at time $t$, while the value zero indicates that the realization $X(\cdot,\omega_i)$ either marginally satisfies $\phi$ at time $t$  or does not satisfy $\phi$ at time $t$. Furthermore, large  positive values of $\text{dist}^{\neg\phi}(X(\cdot,\omega_i),t)$ indicate robust satisfaction and are hence desirable. This is the reason why $-\text{dist}^{\neg\phi}(X,t)$ is considered in Definition \ref{def:rr} as the cost random variable. To complement Fig.~\ref{fig:RD}, note that the red curve sketches a possible distribution of $X$ and hence the probability by which  a realization occurs. Note that the robustness degree of the corresponding realizations in Fig.~\ref{fig:RD}  would be  $\mathcal{RD}^\phi(X(\cdot,\omega_1),t)<0$, $\mathcal{RD}^\phi(X(\cdot,\omega_2),t)= 1$, $\mathcal{RD}^\phi(X(\cdot,\omega_3),t)= 2$, and $\mathcal{RD}^\phi(X(\cdot,\omega_4),t)= 3$.

\begin{example}
	Consider the value-at-risk at level $\beta:=0.95$ (see Appendix C), which is also known as the $1-\beta$ risk quantile. Assume that we obtained $VaR_\beta(-\text{dist}^{\neg\phi}(X,t))=-1$ for a given stochastic process $X$ and a given STL formula $\phi$. The interpretation is now that with a probability of $0.05$ the robustness $\text{dist}^{\neg\phi}(X,t)$ is smaller than (or equal to) $|VaR_\beta(-\text{dist}^{\neg\phi}(X,t))|=1$. Or in other words, with a probability of  $0.95$, the robustness $\text{dist}^{\neg\phi}(X,t)$ is greater than $|VaR_\beta(-\text{dist}^{\neg\phi}(X,t))|=1$.
\end{example}

\begin{remark}
	An alternative to Definition \ref{def:rr} would be to use the robustness degree $\mathcal{RD}^\phi(X,t)$ and to consider $R(-\mathcal{RD}^\phi(X,t))$ instead of $R(-\text{dist}^{\neg\phi}(X,t))$. We, however, refrain from such a definition since the meaning of $\mathcal{RD}^\phi(X(\cdot,\omega_i),t)$ for realizations $w_i\in\Omega$ with $\mathcal{RD}^\phi(X(\cdot,\omega_i),t)<0$ is not what we aim for here. In particular, when $\mathcal{RD}^\phi(X(\cdot,\omega_i),t)<0$ we have that  $\mathcal{RD}^\phi(X(\cdot,\omega_i),t)=-\text{dist}^{\phi}(X,t)$. In this case, $|\mathcal{RD}^\phi(X(\cdot,\omega_i),t)|$ indicates the robustness by which $X(\cdot,\omega_i)$ satisfies $\neg\phi$, while we are interested in the opposite.  
\end{remark}

Unfortunately, the risk of not satisfying $\phi$ robustly, i.e., $R(-\text{dist}^{\neg\phi}(X,t))$, can in most of the cases not be computed. Recall that this was similarly the case for the robustness degree in Definition~\ref{def:rd_cont}. Instead, we will focus on $R(-\rho^\phi(X,t))$ as an approximate risk of not satisfying $\phi$ robustly, i.e., the \emph{approximate STL robustness risk}. We next show that this approximation has the desirable property of being an over-approximation. 

\subsubsection{Approximating the risk of not satisfying $\phi$ robustly} A desirable property is that $R(-\rho^\phi(X,t))$ over-approximates $R(-\text{dist}^{\neg\phi}(X,t))$ so that $R(-\rho^\phi(X,t))$ is more risk-aware than $R(-\text{dist}^{\neg\phi}(X,t))$, i.e., that it holds that $R(-\text{dist}^{\neg\phi}(X,t))\le R(-\rho^\phi(X,t))$. We next show that this property holds when $R$ is monotone.
\begin{theorem}\label{thm:3}
	Let $R$ be a monotone risk measure. Then it holds that $R(-\text{dist}^{\neg\phi}(X,t))\le R(-\rho^\phi(X,t))$.
\end{theorem}

This indeed enables us to use $R(-\rho^\phi(X,t))$ instead of $R(-\text{dist}^{\neg\phi}(X,t))$. In the next section, we elaborate on a data-driven method to estimate  $R(-\rho^\phi(X,t))$ when $R(\cdot)$ is the value-at-risk. For two stochastic processes $X_1$ and $X_2$, note that $R(-\rho^\phi(X_1,t))\le R(-\rho^\phi(X_2,t))$ means that $X_1$ has less risk than $X_2$ with respect to the specification $\phi$.

%\textcolor{red}{Insert some straightforward corollaries for comparison of two stochastic processes with different risk values, put in the running example.}

Oftentimes, one may be interested in associating a monetary cost with  $\text{dist}^{\neg\phi}(X,t)$ that reflects the severity of an event with low robustness. One may hence want to assign high costs to low robustness and low costs to high robustness. Let us define an increasing cost function $C:\mathbb{R}\to\mathbb{R}$ that reflects this preference.
\begin{corollary}
	Let $R$ be a monotone risk measure and $C$ be an increasing cost function. Then it holds that $R(C(-\text{dist}^{\neg\phi}(X,t)))\le R(C(-\rho^\phi(X,t)))$.
\end{corollary}
%The property in Theorem \ref{thm:3}  is indeed useful to upper bound the risk as per Alternative 2. 

%A natural question is, can we obtain a similar relationship between $R(-\mathcal{RD}^\phi(X,t))$ (Alternative 1) and $R(-\rho^\phi(X,t))$. Unfortunately,  the answer is in general no for the simple reason that $-\text{dist}^{\phi}(X(\cdot,\omega),t)\le \rho^\phi(X(\cdot,\omega),t)$ for all $\omega\in\Omega$ as opposed to $\rho^\phi(X(\cdot,\omega),t)\le \text{dist}^{\neg\phi}(X(\cdot,\omega),t)$. For an illustration and more intuition, see also \cite[Example 18]{fainekos2009robustness}.

%\subsection{The Temporal Risk of Violating a Specification}

%The idea here is to define temporal risk. The authors in \cite{donze2} have defined temporal robustness as opposed to spatial robustness in Definition \ref{def:quantitative_semantics}. Let us denote temporal robustness by $\tau^\phi(x,t)$ for a signal $x:T\to\mathbb{R}^n$. We can then define the risk of the stochastic process $X$ not satisyfing the specification $\phi$ at time $t$ as $R(-\tau^\phi(x,t))$.

%\subsection{Vanishing Uncertainty}
%TO BE DONE:  We would now like to analyze that, as the uncertainty of $X$ in form of the covariance matrix goes to zero, we recover the original semantics. This is more like a sanity check.

\section{Data-Driven Estimation of the STL Robustness Risk}
\label{comppp}

There are  two main challenges in computing the approximate STL robustness risk.  First,  note that exact calculation of $R(-\rho^\phi(X,t))$ requires knowledge of the CDF of $\rho^\phi(X,t)$ no matter what the choice of the risk measure $R$ will be.  However, the CDF of $\rho^\phi(X,t)$ is not known (only the CDF of  $X$ is known) and deriving the CDF of $\rho^\phi(X,t)$ is often not possible. Second, calculating $R(-\rho^\phi(X,t))$ may involve solving high dimensional integrals.

In this paper, we assume a data-driven setting where realizations $X^i$ of the stochastic process $X$ are observed, e.g., obtained from experiments, and where not even the CDF of $X$ is known. We  present a data-driven sample average approximation $\overline{R}(-\rho^\phi(X^i,t))$ of $R(-\rho^\phi(X,t))$ for the value-at-risk and show that this approximation has the favorable property of being an upper bound to $R(-\rho^\phi(X,t))$, i.e., that  $R(-\rho^\phi(X,t))\le \overline{R}(-\rho^\phi(X^i,t))$ with high probability. 

\subsection{Sample Average Approximation of the Value-at-Risk (VaR)} Let us first obtain a sample approximation of the value-at-risk (VaR) and define for convenience the random variable 
\begin{align*}
    Z:=-\rho^\phi(X,t).
\end{align*}
For further convenience, let us define the tuple
\begin{align*}
    \mathcal{Z}:=(Z^1,\hdots,Z^N)
\end{align*}
where $Z^i:=-\rho^\phi(X^i,t)$ and where $X^1,\hdots,X^N$ are $N$ independent copies of $X$. Consequently, all $Z^i$ are independent and identically distributed. 

For a risk level of $\beta\in(0,1)$, the VaR of $Z$ is given by
\begin{align*}
VaR_\beta(Z):= \inf\{\alpha\in\mathbb{R}|F_{Z}(\alpha)\ge \beta\}
\end{align*}
where we recall that $F_{Z}$ is the CDF of $Z$. To estimate $F_{Z}(\alpha)$, define the empirical CDF 
\begin{align*}
\widehat{F}(\alpha,\mathcal{Z}):=\frac{1}{N}\sum_{i=1}^N \mathbb{I}(Z^i\le \alpha)
\end{align*}
where we recall that $\mathbb{I}$ denotes the indicator function. Let $\delta\in(0,1)$ be a probability threshold. Inspired by \cite{szorenyi2015qualitative} and \cite{massart1990tight}, we calculate an upper bound of $VaR_\beta(Z)$ as
\begin{align}\label{eq:upper1}
\begin{split}
\overline{VaR}_\beta(\mathcal{Z},\delta)&:=\inf\Big\{\alpha\in \overline{\mathbb{R}}|\widehat{F}(\alpha,\mathcal{Z})-\sqrt{\frac{\ln(2/\delta)}{2N}}\ge \beta\Big\}
\end{split}
\end{align}
and a lower bound as
\begin{align*}
\underline{VaR}_\beta(\mathcal{Z},\delta)&:=\inf\Big\{\alpha\in \overline{\mathbb{R}}|\widehat{F}(\alpha,\mathcal{Z})+\sqrt{\frac{\ln(2/\delta)}{2N}}\ge \beta\Big\}
\end{align*}
where we recall that $\inf \emptyset=\infty$, for $\emptyset$ being the empty set, due to  the extended definition of the infimum operator. We next show that $\overline{VaR}_\beta(\mathcal{Z},\delta)$ is an upper bound of $VaR_\beta(Z)$ with a probability of at least $1-\delta$.

\begin{theorem}\label{thm:mmm_}
     Let $\delta\in(0,1)$ be a probability threshold and $\beta\in(0,1)$ be a risk level. Let $\overline{VaR}_\beta(\mathcal{Z},\delta)$ and $\underline{VaR}_\beta(\mathcal{Z},\delta)$ be based on $\mathcal{Z}$ that are $N$ independent copies of $Z$. Assume that the distribution $F_Z$ is  continuous.  With probability of at least $1-\delta$, it holds that
	\begin{align*}
	\underline{VaR}_\beta(\mathcal{Z},\delta)\le VaR_\beta(Z)\le \overline{VaR}_\beta(\mathcal{Z},\delta).
	\end{align*}
\end{theorem}

Theorem \ref{thm:mmm_} now provides an upper bound $\overline{VaR}_\beta(\mathcal{Z},\delta)$ for the risk $VaR_\beta(Z)$ as we desired, while the lower bound $\underline{VaR}_\beta(\mathcal{Z},\delta)$ indicates how conservative our upper bound might have been. We remark that Theorem \ref{thm:mmm_} assumes that the distribution $F_Z$ is continuous. If $F_Z$ is not continuous, one can derive upper and lower bounds by using order statistics following  \cite[Lemma 3]{nikolakakis2021quantile}.

As a result of Theorems \ref{thm:3} and \ref{thm:mmm_} and the fact that the VaR is a monotone risk measure, we have now a procedure to find a tight upper bound $\overline{VaR}_\beta(\mathcal{Z},\delta)$ of $VaR_\beta(Z)$ with high probability. To summarize, we observe $N$ realizations $X^1,\hdots,X^N$ of the stochastic process $X$. We then select $\delta,\beta\in(0,1)$ and are guaranteed that with probability of at least $1-\delta$ 
\begin{align*}
    VaR_\beta(-\text{dist}^{\neg\phi}(X,t))\le VaR_\beta(Z)\le \overline{VaR}_\beta(\mathcal{Z},\delta).
\end{align*}

\begin{remark}
Upper and lower bounds for other risk measures than the value-at-risk can often be derived. For the expected value $E(-\rho^\phi(X,t))$, concentration inequalities for the sample average approximation of $E(-\rho^\phi(X,t))$ can  be obtained by applying Hoeffding's inequality when $\rho^\phi(X,t)$ is bounded. For the conditional value-at-risk $CVaR(-\rho^\phi(X,t))$,  concentration inequalities are presented in \cite{bhat2019concentration,brown2007large,thomas2019concentration,mhammedi2020pac}. We plan to address this in future work.
\end{remark}

\section{Case Study}
\label{sec:simulations}
We continue with the case study presented in Example \ref{ex1}. Now, however, the environment is uncertain as the regions $C$ and $D$ in which humans operate are not exactly known. Let therefore  $c$ and $d$ be Gaussian random vectors as
\begin{align*}
    c\sim\mathcal{N}\Big(\begin{bmatrix}2 \\ 3 \end{bmatrix},\begin{bmatrix}0.125 & 0\\ 0 & 0.125\end{bmatrix}\Big),\\
    d\sim \mathcal{N}\Big(\begin{bmatrix}6 \\ 4 \end{bmatrix},\begin{bmatrix}0.125 & 0\\ 0 & 0.125\end{bmatrix}\Big),
\end{align*}
where $\mathcal{N}$ denotes a multivariable Gaussian distribution  with according mean vector and covariance matrix. Consequently, the signals $x_1$-$x_6$ become stochastic processes denoted by $X_1$-$X_6$. For each $j\in\{1,\hdots,6\}$ and for $\mathcal{Z}_j:=(-\rho^\phi(X_j^1,t),\hdots,-\rho^\phi(X_j^N,t))$, our goal is now to calculate 
\begin{align*}
    \overline{VaR}_\beta(\mathcal{Z}_j,\delta)
\end{align*}
to compare the risk between the six robot trajectories $r_1$-$r_6$. We set $\delta:=0.001$ and $N:=6500$.  For different $\beta$, the resulting $\overline{VaR}_\beta$ are shown in the following table.
\begin{center}
\begin{tabular}{|p{1.2cm}|p{1.2cm}|p{1.2cm}|p{1.2cm}|p{1.2cm}|}
\hline
\backslashbox{$j$}{$\beta$} & $0.9$ & $0.925$ & $0.95$ & $0.975$ \\
\hline
1 & 0.336 & 0.363 & 0.395 & 0.539 \\
2 & 0.162 & 0.187 & 0.220 & 0.336 \\
3 & -0.17 & -0.152 & -0.121 & -0.008 \\
4 & -0.249 & -0.249 & -0.249 & -0.19\\
5 & -0.25 & -0.25 & -0.25 & -0.149 \\
6 & -0.249 & -0.249 & -0.249 & -0.249 \\
\hline
\end{tabular}
\end{center}

Across all $\beta$, the table indicates that trajectories $r_1$ and $r_2$ are not favorable in terms of the induced STL robustness risk. Trajectory $r_3$ is better compared to trajectories $r_1$ and $r_2$, but worse than $r_4$-$r_6$ in terms of the robustness risk of $\phi$. For trajectories $r_4$-$r_6$, note that a $\beta$ of $0.9$, $0.925$, and $0.95$ provides the information that the trajectories have roughly the same robustness risk. However, once the risk level $\beta$ is increased to $0.975$, it becomes clear that $r_6$ is preferable over $r_4$ that is again preferable over  $r_5$. This matches with what one would expect by closer inspection of Fig. \ref{ex:1_figure}.
\section{Conclusion}
\label{sec:conclusion}
We defined the risk of a stochastic process not satisfying a signal temporal logic specification robustly which we referred to as the ``STL robustness risk''. We also presented an approximation of the STL robustness risk that is an upper bound of the STL robustness risk when the used risk measure is monotone. For the case of the value-at-risk, we presented a data-driven method to estimate the approximate STL robustness risk.

\section*{Acknowledgment}

The authors would like to thank Al\"ena Rodionova and Matthew Cleaveland for proofreading parts of this paper.

\bibliographystyle{IEEEtran}
\bibliography{literature}

\appendix
\section{Semantics and Robust Semantics of STL}
\label{app:STL}
The semantics that are associated with an STL formula $\phi$ as defined in Section \ref{sec:STL} are defined as follows.
\begin{definition}[STL Semantics]\label{def:qualitative_semantics}
For a signal $x:T\to\mathbb{R}^n$, the semantics $\beta^\phi(x,t)$ of an STL formula $\phi$ are inductively defined as
	\begin{align*}
	\beta^\top(x,t)&:=\top,  \\
	\beta^\mu(x,t)&:=\begin{cases}
	\top &\text{ if }	x(t)\in O^\mu\\	
	\bot &\text{ otherwise, }	
	\end{cases}\\
	\beta^{\neg\phi}(x,t)&:= \neg \beta^{\phi}(x,t),\\
	\beta^{\phi' \wedge \phi''}(x,t)&:=\min(\beta^{\phi'}(x,t),\beta^{\phi''}(x,t)),\\
	%	(x,t) \models \phi' \vee \phi'' &\text{ iff } (x,t) \models \phi' \vee (x,t) \models \phi''\\
	\beta^{\phi' U_I \phi''}(x,t)&:=\sup_{t''\in (t\oplus I)\cap T}\Big( \min\big(\beta^{\phi''}(x,t''),\inf_{t'\in(t,t'')\cap T}\beta^{\phi'}(x,t')\big)\Big),\\
	\beta^{\phi' \underline{U}_I \phi''}(x,t)&:=\sup_{t''\in (t\ominus I)\cap T}\Big( \min\big(\beta^{\phi''}(x,t''),\inf_{t'\in(t,t'')\cap T}\beta^{\phi'}(x,t')\big)\Big).
	\end{align*}
\end{definition}

\section{Random Variables}
\label{app:ran_var}
We can associate a probability space $(\mathbb{R}^n,\mathcal{B}^n,P_Z)$ with the random vector $Z$ where, for Borel sets $B\in\mathcal{B}^n$, the probability measure $P_Z:\mathcal{B}^n\to[0,1]$ is defined as 
\begin{align*}
P_Z(B):=P(Z^{-1}(B))
\end{align*} 
where $Z^{-1}(B):=\{\omega\in\Omega|Z(w)\in B\}$ is the inverse image of $B$ under $Z$.\footnote{Measurability of $Z$ ensures that, for $B\in\mathcal{B}$, $Z^{-1}(B)\in\mathcal{F}$ so that the probability measure $P$ can be pushed through to obtain $P_Z$. } In particular $P_Z$ now describes the \emph{distribution} of $Z$. For vectors $z\in\mathbb{R}^n$,  the \emph{cumulative distribution function (CDF)} of $Z$ is  defined as 
\begin{align*}
F_Z(z)=P_Z((-\infty,z_1]\times\hdots\times(-\infty,z_n])
\end{align*} 
where $z_i$ is the $i$th element of $z$. When the CDF $F_Z(z)$ is absolutely continuous, i.e., when $F_Z(z)$ can be written as
\begin{align*}
F_Z(z)=\int_{-\infty}^z f_Z(z') \mathrm{d}z'
\end{align*} 
for some non-negative and Lebesgue measurable function $f_Z(z)$, then $Z$ is a \emph{continuous random vector} and $f_Z(z)$ is called the \emph{probability density function (PDF)} of $Z$.
When the CDF $F_Z(z)$, on the other hand, is discontinuous, then $Z$ is a \emph{discrete random vector}\footnote{There also exist mixed random variables, i.e., random variables with  continuous and discrete parts, which we do not cover explicitly for simplicity and without  loss of generality.} and $f_Z(z)$ is called the \emph{probability mass function (PMF)} satisfying
\begin{align*}
F_Z(z)=\sum_{z'\le z}f_Z(z').
\end{align*} 
The results that we present in this paper apply to both continuous and discrete random variables.\footnote{Note that there exists CDFs that are neither absolutely continuous nor discontinuous, e.g., the Cantor distribution, and have no PDF and no PMF. Our results also apply to these distributions.}

\section{Risk Measures}
\label{app:risk}
We next present some desireable properties that a risk measure may have. Let therefore $Z,Z'\in \mathfrak{F}(\Omega,\mathbb{R})$ be cost random variables. A risk measure is  \emph{coherent} if the following four properties are satisfied.\\
\emph{1. Monotonicity:} If $Z(\omega) \leq Z'(\omega)$ for all $\omega\in\Omega$, it holds that $R(Z) \le R(Z')$.\\
\emph{2. Translation Invariance:} Let $c\in\mathbb{R}$. It holds that $R(Z + c) = R(Z) + c$.\\
\emph{3. Positive Homogeneity:} Let $c\in\mathbb{R}_{\ge 0}$. It holds that $R(c Z) =  cR(Z)$.\\
\emph{4. Subadditivity:} It holds that $R(Z + Z') \leq R(Z) + R(Z')$.

If the risk measure additionally satisfies the following two properties, then it is called a distortion risk measure. 

\noindent \emph{5. Comonotone Additivity:} If $(Z(\omega) - Z(\omega'))(Z'(\omega) - Z'(\omega')) \ge 0$ for all $\omega, \omega' \in \Omega$ (namely, $Z$ and $Z'$ are commotone), it holds that $R(Z + Z') = R(Z) + R(Z')$.\\
\emph{6. Law Invariance:} If $Z$ and $Z'$ are identically distributed, then $R(Z) = R(Z')$.

Common examples of popular risk measures are the expected value $\text{E}(Z)$ (risk neutral) and the worst-case $\text{ess} \sup_{\omega\in\Omega} Z(\omega)$ as well as:
\begin{itemize}
	\item Mean-Variance: $\text{E}(Z) + \lambda \text{Var}(Z)$ where  $\lambda> 0$.
	\item Value at Risk (VaR) at level $\beta \in (0,1)$: 
	\begin{align*}
	    VaR_\beta(Z):=\inf \{ \alpha \in \mathbb{R} |  F_Z(\alpha) \ge \beta \}.
	\end{align*}
	\item Conditional Value at Risk (CVaR) at level $\beta \in (0,1)$: 
	\begin{align*}
	    CVaR_\beta(Z):=\inf_{\alpha \in \mathbb{R}} \; \alpha+(1-\beta)^{-1}E([Z-\alpha]^+)
\end{align*} 
	where $[Z-\alpha]^+:=\max(Z-\alpha,0)$. When the CDF $F_Z$ of $Z$ is continuous, it holds that $CVaR_\beta(Z):=E(Z|Z\ge VaR_\beta(Z))$, i.e., $VaR_\beta(Z)$ is the expected value of $Z$ conditioned on the event that $Z$ is greater than  $VaR_\beta(Z)$.
\end{itemize}

Many risk measures are not coherent and can lead to a misjudgement of risk, e.g., the mean-variance is not monotone and the  value at risk (which is closely related to chance constraints as often used in optimization) does not satisfy the subadditivity property.

\section{Proof of Theorem \ref{thm:1}} \emph{Semantics.} Let us define the power set of  $\mathbb{B}$ as $2^\mathbb{B}:=\{\emptyset,\top,\bot,\{\bot,\top\}\}$. Note that $2^\mathbb{B}$ is a  $\sigma$-algebra of $\mathbb{B}$. To prove measurability of $\beta^\phi(X(\cdot,\omega),t)$ in $\omega$  for a fixed $t\in T$, we need to show that, for each $B\in2^\mathbb{B}$, it holds that the inverse image  of $B$ under $\beta^\phi(X(\cdot,\omega),t)$ for a fixed $t\in T$ is contained within $\mathcal{F}$, i.e., that it holds that 
\begin{align*}
    \{\omega\in\Omega| \beta^\phi(X(\cdot,\omega),t)\in B\}\subseteq\mathcal{F}.
\end{align*} 
We show measurability of $\beta^\phi(X(\cdot,\omega),t)$ in $\omega$ for a fixed $t\in T$ inductively on the structure of $\phi$. 

$\top$: For $B\in2^\mathbb{B}$, it trivially holds that $\{\omega\in\Omega| \beta^\top(X(\cdot,\omega),t)\in B\}\subseteq\mathcal{F}$ since $\beta^\top(X(\cdot,\omega),t)=\top$ for all $\omega\in\Omega$. This follows according to Definition \ref{def:qualitative_semantics} so that 
\begin{align*}
   \{\omega\in\Omega| \beta^\top(X(\cdot,\omega),t)\in B\}=\emptyset\subseteq\mathcal{F} \text{ if } B\in\{\emptyset,\bot\} 
\end{align*}
and similarly 
\begin{align*}
    \{\omega\in\Omega| \beta^\top(X(\cdot,\omega),t)\in B\}=\Omega\subseteq\mathcal{F} \text{ otherwise}. 
\end{align*}

$\mu$: Let  $1_{O^\mu}:\mathbb{R}^n\to\mathbb{B}$ be the characteristic function of $O^\mu$ with $1_{O^\mu}(\zeta):=\top$ if  $\zeta\in O^\mu$ and $1_{O^\mu}(\zeta):=\bot$ otherwise. According to Definition \ref{def:qualitative_semantics}, we can now write $\beta^{\mu}(X(\cdot,\omega),t)=1_{O^\mu}(X(t,\omega))$.  Recall that $O^\mu$ is  measurable and note that the characteristic function of a measurable set is measurable again (see e.g., \cite[Chapter 1.2]{durrett2019probability}). Since $X(t,\omega)$ is measurable in $\omega$ for a fixed $t\in T$ by definition, it  follows that $1_{O^\mu}(X(t,\omega))$ and hence $\beta^{\mu}(X(\cdot,\omega),t)$ is measurable in $\omega$ for a fixed $t\in T$.  In other words, for $B\in2^\mathbb{B}$, it  follows that 
\begin{align*}
    \{\omega\in\Omega| \beta^{\mu}(X(\cdot,\omega),t)\in B\}=\{\omega\in\Omega|1_{O^\mu}(X(t,\omega))\in B\}\subseteq \mathcal{F}. 
\end{align*}

$\neg\phi$: By the induction assumption, $\beta^{\phi}(X(\cdot,\omega),t)$ is measurable in $\omega$  for a fixed $t\in T$. Recall that $\mathcal{F}$ is a $\sigma$-algebra that is, by definition, closed under its complement so that, for $B\in2^\mathbb{B}$, it holds that  
\begin{align*}
    \{\omega\in\Omega| \beta^{\neg\phi}(X(\cdot,\omega),t)\in B\}=\Omega\setminus \{\omega\in\Omega| \beta^{\phi}(X(\cdot,\omega),t)\in B\}\subseteq \mathcal{F}.
\end{align*}

$\phi'\wedge\phi''$: By the induction assumption, $\beta^{\phi'}(X(\cdot,\omega),t)$ and $\beta^{\phi''}(X(\cdot,\omega),t)$ are measurable in $\omega$  for a fixed $t\in T$. Hence $\beta^{\phi'\wedge\phi''}(X(\cdot,\omega),t)=\min(\beta^{\phi'}(X(\cdot,\omega),t),\beta^{\phi''}(X(\cdot,\omega),t))$ is measurable in $\omega$ for a fixed $t\in T$ since the min operator of measurable functions is again a measurable function.

$\phi' U_I \phi''$ and $\phi' \underline{U}_I \phi''$: Recall that 
\begin{align*}
   \beta^{\phi' U_I \phi''}(X(\cdot,\omega),t) := \underset{t''\in (t\oplus I)\cap T}{\text{sup}}  \big(\min(\beta^{\phi''}(X(\cdot,\omega),t''),\underset{t'\in (t,t'')\cap T}{\text{inf}}\beta^{\phi'}(X(\cdot,\omega),t') )\big). 
\end{align*} 
By the induction assumption, $\beta^{\phi'}(X(\cdot,\omega),t)$ and $\beta^{\phi''}(X(\cdot,\omega),t)$ are measurable in $\omega$ for a fixed $t\in T$. First note that $(t,t'')\cap T$ and $(t\oplus I)\cap T$ are countable sets since $T=\mathbb{N}$. According to \cite[Theorem 4.27]{guide2006infinite}, the supremum and infimum operators over a countable number of measurable functions is again measurable.  Consequently, the function  $\beta^{\phi' U_I \phi''}(X(\cdot,\omega),t)$ is measurable in $\omega$ for a fixed $t\in T$. The same reasoning applies to $\beta^{\phi' \underline{U}_I \phi''}(X(\cdot,\omega),t)$. 

\emph{Robust semantics.} The proof for $\rho^\phi(X(\cdot,\omega),t)$ follows again inductively on the structure of $\phi$ and the goal is to show that  $\{\omega\in\Omega| \rho^\phi(X(\cdot,\omega),t)\in B\}\subseteq\mathcal{F}$ for each Borel set $B\in\mathcal{B}$. The difference here, compared to the proof for the semantics $\beta^\phi(X(\cdot,\omega),t)$ presented above, lies only in the way predicates $\mu$ are handled. Note next that, according to \cite[Lemma 57]{fainekos2009robustness}, it holds for predicates $\mu$ that
\begin{align*}
\text{dist}^{\mu}(X(\cdot,\omega),t)&=\bar{\kappa}\big(X(\cdot,\omega),\text{cl}(\mathcal{L}^\mu(t))\big)=\inf_{x^*\in \text{cl}(\mathcal{L}^\mu(t))}\kappa(X(\cdot,\omega),x^*)\\ &=\inf_{x'\in \text{cl}(O^\mu)} d(X(t,\omega),x')=:\bar{d}\big(X(t,\omega),\text{cl}(O^\mu)\big).
\end{align*}  
We hence have that $\text{dist}^{\mu}(X(\cdot,\omega),t)= \bar{d}(X(t,\omega),\text{cl}(O^\mu))$ and $\text{dist}^{\neg\mu}(X(\cdot,\omega),t)= \bar{d}(X(t,\omega),\text{cl}(O^{\neg\mu}))$.  Consequently, $\rho^{\mu}(X(\cdot,\omega),t)$ encodes the signed distance from the point $X(t,\omega)$ to the set $ O^\mu$ so that
\begin{align}\label{eq:rho_mu}
\begin{split}
\rho^\mu(X(\cdot,\omega),t)&=0.5( 1_{O^\mu}(X(t,\omega))+1)\bar{d}(X(t,\omega),\text{cl}(O^{\neg\mu}))+0.5(1_{O^\mu}(X(t,\omega))-1) \bar{d}(X(t,\omega),\text{cl}(O^\mu))
\end{split}
\end{align} 
where we recall that we interpret $\top:=1$ and $\bot=-1$. Since the composition of the characteristic function with $X(t,\omega)$, i.e.,  $1_{O^\mu}(X(t,\omega))$, is  measurable in $\omega$ for a fixed $t\in T$ as argued before, we only need to  show that $ \bar{d}(X(t,\omega),\text{cl}(O^\mu))$ and  $\bar{d}(X(t,\omega),\text{cl}(O^{\neg\mu}))$ are measurable in $\omega$ for a fixed $t\in T$. This immediately follows since $X(t,\omega)$ is measurable in $\omega$ for a fixed $t\in T$ by definition  and since the function $\bar{d}$ is continuous in its first argument, and hence  measurable (see \cite[Corollary 4.26]{guide2006infinite}), due to $d$ being a metric defined on the set $\mathbb{R}^n$ (see e.g., \cite[Chapter 3]{munkres1975prentice}) so that $\rho^\mu(X(\cdot,\omega),t)$ is measurable in $\omega$ for a fixed $t\in T$.
		
\section{Proof of Theorem \ref{thm:2_}}

For $\text{dist}^\phi(X(\cdot,\omega),t)$, note that, for a fixed $t\in T$, the function $\text{dist}^\phi$ maps from the domain $\mathcal{F}(T,\mathbb{R}^n)$ into the domain $\mathbb{R}$, while $X(\cdot,\omega)$ maps from the domain $\Omega$ into the domain $\mathcal{F}(T,\mathbb{R}^n)$. Recall now that $\text{dist}^\phi(X(\cdot,\omega),t)$ was defined like
\begin{align*}
    \text{dist}^\phi(X(\cdot,\omega),t)=\bar{\kappa}\big(X(\cdot,\omega),\text{cl}(\mathcal{L}^\phi(t))\big):=\inf_{x^*\in \text{cl}(\mathcal{L}^\phi(t))}\kappa(X(\cdot,\omega),x^*)
\end{align*} 
and that $\kappa$ is a metric defined on the set $\mathfrak{F}(T,\mathbb{R}^n)$ as argued in \cite{fainekos2009robustness}. Therefore, it follows that the function $\bar{\kappa}$ is continuous in its first argument (see e.g., \cite[Chapter 3]{munkres1975prentice}), and hence measurable with respect to the Borel $\sigma$-algebra of $\mathfrak{F}(T,\mathbb{R}^n)$ (see e.g., \cite[Corollary 4.26]{guide2006infinite}). Consequently, the function $\text{dist}^\phi:\mathfrak{F}(T,\mathbb{R}^n)\times T\to \mathbb{R}^n$  is  measurable in its first argument for a fixed $t\in T$. As $T$ is countable and due to the assumption that $X$ is a discrete-time stochastic process, it follows that $X(\cdot,\omega)$ is measurable with respect to the product $\sigma$-algebra of Borel $\sigma$-algebras $\mathcal{B}^n$ which is equivalent to the Borel $\sigma$-algebra of $\mathfrak{F}(T,\mathbb{R}^n)$ (see e.g., \cite[Lemma 1.2]{kallenberg1997foundations}). Since function composition preserves measurability, it holds that $\text{dist}^\phi(X(\cdot,\omega),t)$ is measurable in $\omega$ for a fixed $t\in T$. 

%The function $X(\cdot,\omega)$ is measurable in $\omega$ even in the case where $T=\mathbb{R}$ (see e.g., \cite[Prop. 2.6]{capasso2005introduction}).
 
For $\mathcal{RD}^\phi(X(\cdot,\omega),t)$, note that we can write
\begin{align*}
\mathcal{RD}^\phi(X(\cdot,\omega),t)&=0.5( 1_{\mathcal{L}^\phi(t)}(X(\cdot,\omega))+1)\text{dist}^{\neg\phi}(X(\cdot,\omega),t)+0.5(1_{\mathcal{L}^\phi(t)}(X(\cdot,\omega))-1) \text{dist}^\phi(X(\cdot,\omega),t)
\end{align*} 
similarly to \eqref{eq:rho_mu}. While we have shown measurability of $\beta^\phi(X(\cdot,\omega),t)$ in $\omega$ for a fixed $t\in T$ in the proof of Theorem~\ref{thm:1},  we can similarly show measurability of $\beta^\phi(X(\cdot,\omega),t)$ in its first argument  with respect to the Borel $\sigma$-algebra of $\mathfrak{F}(T,\mathbb{R}^n)$. It hence holds that $\mathcal{L}^\phi(t)$ is a measurable set.  Consequently, the functions $1_{\mathcal{L}^\phi(t)}(X(\cdot,\omega))$  and  $\text{dist}^\phi(X(\cdot,\omega),t)$ are measurable in $\omega$ for a fixed $t\in T$. It follows, using similar arguments proceeding \eqref{eq:rho_mu}  in the proof of Theorem~\ref{thm:1}, that $\mathcal{RD}^\phi(X(\cdot,\omega),t)$  is measurable in $\omega$ for a fixed $t\in T$.

\section{Proof of Theorem \ref{thm:3}}

First note that $\rho^\phi(X(\cdot,\omega),t)\le \text{dist}^{\neg\phi}(X(\cdot,\omega),t)$ for each realization $X(\cdot,\omega)$ of the stochastic process $X$ with $\omega\in\Omega$ due to \eqref{underapprox}. Consequently, we have that $ -\text{dist}^{\neg\phi}(X(\cdot,\omega),t)\le -\rho^\phi(X(\cdot,\omega),t)$ for all $\omega\in\Omega$.  If $R$ is now monotone, it directly follows that $R(-\text{dist}^{\neg\phi}(X,t))\le R(-\rho^\phi(X,t))$.

\section{Proof of Theorem \ref{thm:mmm_}:}

We first recall the tight version of the Dvoretzky-Kiefer-Wolfowitz inequality as originally presented in~\cite{massart1990tight} which requires that $F_Z$ is continuous.
\begin{lemma}\label{lem:1_}
	Let $\widehat{F}_{Z}(\alpha,\mathcal{Z})$ be the empirical CDF of the random variable $Z$ where  $\mathcal{Z}:=(Z^1,\hdots,Z^N)$ is a tuple of $N$ independent copies of $Z$. Let $c>0$ be a desired precision, then it holds that
	\begin{align*}
	P\big(\sup_\alpha|\widehat{F}(\alpha,\mathcal{Z})-{F}_{Z}(\alpha)|>c\big) \le 2\exp\big( -2 Nc^2\big).
	\end{align*}
\end{lemma}

By setting $\delta:=2\exp\big( -2 Nc^2\big)$ in Lemma \ref{lem:1_}, it holds  with a probability of at least $1-\delta$ that
\begin{align*}
	\sup_\alpha|\widehat{F}(\alpha,\mathcal{Z})-{F}_{Z}(\alpha)| \le \sqrt{\frac{\ln(2/\delta)}{2N}}.
	\end{align*}
With a probability of at least $1-\delta$, it now holds that 
\begin{align*}
&\{\alpha\in \mathbb{R}|\widehat{F}(\alpha,\mathcal{Z})- \sqrt{\frac{\ln(2/\delta)}{2N}}\ge \beta\}
\subseteq \{\alpha\in \mathbb{R}|{F}_{Z}(\alpha)\ge \beta\}.
\end{align*}
as well as
\begin{align*}
&\{\alpha\in \mathbb{R}|\widehat{F}(\alpha,\mathcal{Z})+ \sqrt{\frac{\ln(2/\delta)}{2N}}\ge \beta\}
\supseteq \{\alpha\in \mathbb{R}|{F}_{Z}(\alpha)\ge \beta\}.
\end{align*}
Hence, it holds with a probability of at least $1-\delta$ that
\begin{align*}
\overline{VaR}_\beta(\mathcal{Z},\delta)=&\inf\{\alpha\in \mathbb{R}|\widehat{F}(\alpha,\mathcal{Z})
 - \sqrt{\frac{\ln(2/\delta)}{2N}}\ge \beta\} 
\ge \inf\{\alpha\in \mathbb{R}|{F}_{Z}(\alpha)\ge \beta\}=VaR_\beta(Z)
\end{align*}
as well as 
\begin{align*}
\underline{VaR}_\beta(\mathcal{Z},\delta)=&\inf\{\alpha\in \mathbb{R}|\widehat{F}(\alpha,\mathcal{Z})
 + \sqrt{\frac{\ln(2/\delta)}{2N}}\ge \beta\}
 \le \inf\{\alpha\in \mathbb{R}|{F}_{Z}(\alpha)\ge \beta\}=VaR_\beta(Z).
\end{align*}
by definition of $VaR_\beta(Z)$, $\underline{VaR}_\beta(\mathcal{Z},\delta)$, and $\overline{VaR}_\beta(\mathcal{Z},\delta)$. In summary, it holds with a probability of at least $1-\delta$ that
\begin{align*}
    \underline{VaR}_\beta(\mathcal{Z},\delta)\le VaR_\beta(Z)\le \overline{VaR}_\beta(\mathcal{Z},\delta).
\end{align*}

\addtolength{\textheight}{-12cm}   % This command serves to balance the column lengths

\end{document}